\documentclass[showpacs,showkeys,superscriptaddress,aps]{revtex4}
\begin{document}
\title{Hyper-elliptic Nambu flow associated with integrable maps}
\author{Satoru SAITO}
\email[email : ]{saito@phys.metro-u.ac.jp}
\affiliation{Department of Physics, Tokyo Metropolitan University,\\
Minamiohsawa 1-1, Hachiohji, Tokyo 192-0397 Japan}
\author{Noriko SAITOH}
\email[email : ]{saitoh@lam.osu.sci.ynu.ac.jp}
\affiliation{Department of Applied Mathematics, Yokohama National University\\
Hodogaya-ku, Yokohama, 240-8501 Japan}
\author{Katsuhiko YOSHIDA}
\email[email : ]{yoshida@kiso.phys.metro-u.ac.jp}
\affiliation{School of Science, Kitasato University,\\ 1-15-1 Kitasato Sagamihara, Kanagawa, 228-8555 Japan}
\begin{abstract}
We study hyper-elliptic Nambu flows associated with some $n$ dimensional maps and show that discrete integrable systems can be reproduced as flows of this class.
\end{abstract}
\maketitle
\baselineskip 20pt
\section{Introduction}

Let us consider an $n$ dimensional rectangular box with edges of length $X_1,X_2,\cdots, X_n$. The box will be rigid if there are $n$ independent relations among $X_j$'s
\begin{equation}
x_j=f_j(X_1,X_2,\cdots, X_n),\qquad j=1,2,\cdots, n.
\label{relations}
\end{equation}
This set of constraints defines a map
\begin{equation}
\mbox{\boldmath$x$}=(x_1,x_2,\cdots, x_n)\rightarrow \mbox{\boldmath$X$}=(X_1,X_2,\cdots, X_n).
\label{map}
\end{equation}

If we relax one of the constraints, say $x_n=f_n$, the box changes its form as $x_n$ varies. We are interested in how it changes. To make clear the problem let us see the case of $n=3$ and assume, for example, that the relations (\ref{relations}) are given by the elementary symmetric polynomials
\begin{eqnarray}
x_1&=&X_1+X_2 +X_3\nonumber\\
x_2&=&X_1X_2+X_1X_3+X_2X_3
\label{3 symmetric poly}\\
x_3&=&X_1X_2X_3.\nonumber
\end{eqnarray}

When all $x_j$'s are fixed the map is determined algebraically by solving the equation
\begin{equation}
X^3-x_1X^{2}+x_2X-x_3=0,
\end{equation}
up to permutations. The relations (\ref{3 symmetric poly}) amount to fix the total length of edges, the total area of surfaces and the volume of the box. If the volume of the box $x_3$ varies while $x_1$ and $x_2$ are fixed, we will obtain a circle as an intersection of the sphere $X_1^2+X_2^2+X_3^2=x_1^2-2x_2$ and the plane $X_1+X_2+X_3=x_1$ in $\mbox{\boldmath$R$}^3$ along which \mbox{\boldmath$X$} moves. If we fix $x_1$ and $x_3$ but leave $x_2$ free, we will find another curve in $\mbox{\boldmath$R$}^3$. Our problem is to find such a curve in general.

It will be worthwhile to notice here that the relations (\ref{relations}) defines an algebraic manifold in $\mbox{\boldmath$R$}^n$ if some of $x_j$'s are fixed and the relations are purely algebraic. To see the properties of the manifold one can leave one of the constraints free and vary the constant to study a response of the variables $X_1,X_2,\cdots,X_n$ to the variation.

In our previous work\cite{SSYY,SSYY2} we have shown that there exists a Nambu-Hamiltonian flow\cite{Nambu} corresponding to an arbitrary differentiable map such that one of the initial values of the map plays the role of time of the flow. We can apply this result to see the change of the box when $x_n$ varies.

We will study, in this article, how the problem is transcribed into the problem of solving the Nambu equation and the change of the boxes are described in terms of hyper-elliptic curves under certain constraints. We are interested in the appearance of the hyper-elliptic functions, since they are known to solve soliton equations in general. The second purpose of this article is to show that the reason of the appearance of the hyper-elliptic functions is common in two systems, {\it i.e.}, the soliton equations and the Nambu equations. In other words the Nambu equations provide a way to describe dynamics of integrable systems. We will show that, once the Nambu equations associated with a soliton equation are solved, solutions to the soliton equation can be given by solving purely algebraic relations.

\section{Nambu-Hamiltonian flows}

Let us first recall briefly the Nambu equations and review our previous results. For a function $f(\mbox{\boldmath$X$})$ of $n$-dimensional variable $\mbox{\boldmath$X$}\in \mbox{\boldmath$R$}^n$, the generalized Nambu Hamilton equations\cite{Nambu,Takhtajan} are given by
\begin{equation}
{df\over dt}(\mbox{\boldmath$X$})=\{H_1,H_2,\cdots,H_{n-1},f(\mbox{\boldmath$X$})\}.
\label{Nambu eq for f}
\end{equation}
We define the Nambu bracket $\{\varphi_1,\varphi_2,\cdots, \varphi_n\}$, in this article, by the Jacobian
$$
\{\varphi_1,\varphi_2,\cdots, \varphi_n\}:={\partial(\varphi_1,\varphi_2,\cdots, \varphi_n)\over \partial(X_1, X_2, \cdots, X_n)}.
$$
$n-1$ Hamiltonians $H_1,H_2,\cdots, H_{n-1}$ satisfy
$$
{dH_j\over dt}=0,\qquad j=1,2,\cdots, n-1
$$
by definition of the equations. The equations of motion for the dynamical variables $X_j$'s are
\begin{eqnarray}
{dX_j\over dt}&=&\{H_1,H_2,\cdots,H_{n-1},X_j\}, \qquad j=1,2,\cdots, n
\nonumber\\
&=&(-1)^{n-j}{\partial(H_1,H_2,\cdots, H_{n-1})\over \partial(X_1, X_2, \cdots, X_n)_j}.
\label{Nambu eq}
\end{eqnarray}
Here $(X_1, X_2, \cdots, X_n)_j$ means that $X_j$ is missing among $(X_1, X_2, \cdots, X_n)$.

In \cite{SSYY,SSYY2} we proved the following:
\vglue 0.5cm
\noindent
{\bf Proposition 1}

{\it For a differentiable and invertible map (\ref{map}), with its Jacobian $\det J$, there exists a Nambu-Hamiltonian flow described by the equations
}
\begin{equation}
{dX_j\over dx_n}=\{H_1,H_2,\cdots, H_{n-1},X_j\},\qquad j=1,2,\cdots, n,
\label{dX_j/dx_n}
\end{equation}
{\it such that the Hamiltonians are given by}
\begin{eqnarray*}
H_j&=&x_j,\qquad j=1,2,\cdots, n-2,\\
H_{n-1}&=&\int^{x_{n-1}}(\det J)\ dx_{n-1}.
\end{eqnarray*}
\vglue 0.3cm

Note that the initial value $x_n$ of the map plays the role of the time variable in this flow. In addition to this proposition we would like to supply a new one which is more convenient to study our present problems.

\vglue 0.5cm
\noindent
{\bf Proposition 2}

{\it For a differentiable and invertible map (\ref{map}), with its Jacobian $\det J$, there exists a Nambu-Hamiltonian flow described by the equations
\begin{equation}
{dX_j\over dt}=\{H_1,H_2,\cdots, H_{n-1},X_j\},\qquad j=1,2,\cdots, n,
\label{dX_j/dt}
\end{equation}
such that the Hamiltonians are given by
$$
H_j=x_j,\qquad j=1,2,\cdots, n-1,
$$
and the variable $t$ satisfies
\begin{equation}
{dx_k\over dt}={\delta_{k,n}\over\det J}.
\label{t=}
\end{equation}
}
\vglue 0.5cm

Here the time variable $t$ is not the initial value $x_n$ itself but related to it by (\ref{t=}). This formulation has an advantage since the other initial values $x_1,x_2,\cdots, x_{n-1}$ of the map remain constant along the Nambu flow.

The proof of Proposition 2 is straightforward. From (\ref{t=}) it follows that 
\begin{eqnarray*}
{dX_j\over dt }&=&
\sum_k{\partial X_j\over\partial x_k}{dx_k\over dt}
=
{\partial X_j\over\partial x_n}{1\over \det J}\\
&=&
{\partial(x_1,x_2,\cdots, x_{n-1},X_j)\over\partial(x_1,x_2,\cdots, x_{n-1},x_n)}{1\over\det J}\\
&=&
{\partial(x_1,x_2,\cdots, x_{n-1},X_j)\over\partial(X_1,X_2,\cdots, X_{n-1},X_n)}\\
&=&
{\partial(H_1,H_2,\cdots, H_{n-1},X_j)\over\partial(X_1,X_2,\cdots, X_{n-1},X_n)}.\\
&&\qquad\qquad\qquad\qquad\qquad\qquad\qquad {\it q.e.d.}
\end{eqnarray*}

Hence (\ref{dX_j/dt}) is true. Conversely (\ref{dX_j/dt}) implies $dH_j/dt=0$, hence (\ref{t=}) follows. We notice that, when the Nambu equation (\ref{dX_j/dt}) is solved,
\begin{equation}
t=\int^{x_n}(\det J)\ dx_n
\label{t=int det J}
\end{equation}
holds. If the Jacobian $\det J$ of the map $\mbox{\boldmath$x$}\rightarrow \mbox{\boldmath$X$}$ was one, we simply have $t=x_n$.

Now suppose we have solved the Nambu equations (\ref{dX_j/dt}). We then obtain a map $(x_1,\cdots,x_{n-1},t)\rightarrow \mbox{\boldmath$X$}$. We can show that the Jacobian $\det J'$ of this map is unity. 

To see that we calculate the Jacobian of the inverse map $\mbox{\boldmath$X$} \rightarrow (x_1,\cdots,x_{n-1},t)$ and expand it along the last row
$$
\det J'^{-1}={\partial(x_1,x_2,\cdots, x_{n-1},t)\over \partial(X_1,X_2,\cdots, X_n)}=
\sum_{j=1}^n{\partial t\over \partial X_j}\Delta_{nj}
$$
where $\Delta_{nj}$ is the minor of the $(nj)$ element of $J'^{-1}$. We notice that the right hand side of (\ref{dX_j/dt}) is exactly $\Delta_{nj}$. Therefore we obtain
$$
\det J'^{-1}=\sum_{j=1}^n{\partial t\over \partial X_j}{dX_j\over dt}=1
$$
as a result of equations of motion. Therefore the Jacobian $\det J'$ of the map $(x_1,\cdots,x_{n-1},t) \rightarrow \mbox{\boldmath$X$}$ is also one and the map preserves the hyper-volume element.

\section{Study of solutions}

Since $x_j, \ j=1,2,\cdots, n-1$ are constants of the flow (\ref{dX_j/dt}), they form $n-1$ dimensional hypersurfaces
\begin{equation}
x_j=H_j(\mbox{\boldmath$X$}), \qquad j=1,2,\cdots, n-1
\label{constraints}
\end{equation}
in $\mbox{\boldmath$R$}^n$, while the point $\mbox{\boldmath$X$}$ moves along a curve formed by an intersection of the hypersurfaces as $t$ changes. Suppose we can solve the constraints (\ref{constraints}) for $X_1,X_2,\cdots, X_{n-1}$ as functions of $W:=X_n$ and the constants $x_j,\ j=1,2,\cdots, n-1$. Substituting the results into the right hand side of the equation for $W$ in (\ref{dX_j/dt}), we obtain a first order ordinary differntial equation for $W$,
\begin{equation}
{dW\over dt}=F(W),
\label{dW/dt=F}
\end{equation}
where $F$ is a function of $W$ and the constants of the flow. The orbit is determined by solving (\ref{dW/dt=F}), {\it i.e.},
\begin{equation}
t=\int^{W}{dW\over F(W)}.
\label{t=int dW/F}
\end{equation}

Combining this result with (\ref{t=int det J}) we find
$$
{dW\over dx_n}=F(W)\det J.
$$
Similarly we obtain equations for all other variables $X_j$'s which determine the dependence on $x_n$
.

\subsection{Elementary symmetric polynomials}

First we study a Nambu flow when the constraints (\ref{relations}) are given by the elementary symmetric polynomials
\begin{eqnarray}
x_1&=&X_1+X_2 +X_3+\cdots +X_n\nonumber\\
x_2&=&X_1X_2+X_1X_3+X_2X_3+\cdots +X_{n-1}X_n\nonumber\\
&\vdots&\nonumber\\
x_{j}&=&\sum_{k_1<k_2<\cdots<k_j}X_{k_1}X_{k_2}\cdots X_{k_j}\label{symmetric poly}\\
&\vdots&\nonumber\\
x_n&=&X_1X_2\cdots X_n.\nonumber
\end{eqnarray}

When all $x_j$'s are fixed the map is determined by solving the algebraic equation
\begin{equation}
X^n-x_1X^{n-1}+\cdots -(-1)^nx_{n-1}X+(-1)^nx_n=0,
\end{equation}
up to permutations. The relations (\ref{symmetric poly}) amount to fix the total length of edges, the total area of surfaces, $\cdots$, the total hyper-volume of the box. If the hyper-volume of the box $x_n$ varies while other $x_j$'s are fixed, we will obtain a curve in $\mbox{\boldmath$R$}^n$ along which \mbox{\boldmath$X$} moves. Our problem is to find the curve.

The Nambu equation whose Hamiltonians are $x_1,x_2,\cdots, x_{n-1}$ in (\ref{symmetric poly}) is given for $W=X_n$ \cite{Takhtajan} by
$$
{dW\over dt}=\prod_{1\le k<l\le n-1}(X_k-X_l).
$$
Therefore our task is to solve this equation explicitly. For this to be done we have to know the $W$ dependence of the right hand side. We first notice that the square of the right hand side is the discriminant of the equation $P_{n-1}(X)=0$, where
\begin{equation}
P_{n-1}(X):=(X-X_1)(X-X_2)\cdots (X-X_{n-1}).
\end{equation}
If we expand the polynomial $P_{n-1}(X)$ as
\begin{eqnarray*}
P_{n-1}(X)&=&h_0X^{n-1}-h_1X^{n-2}+h_2X^{n-3}\\
&&\qquad -\cdots +(-1)^{n-1}h_{n-1},\quad (h_0=1)
\end{eqnarray*}
$h_1,h_2,\cdots, h_{n-1}$ are the elementary symmetric polynomials of $X_1,X_2,\cdots, X_{n-1}$. Since the discriminant
$$
D_{n-1}:=\prod_{1\le k<l\le n-1}(X_k-X_l)^2
$$
of $P_{n-1}(X)=0$ is a symmetric polynomial it can be expressed in terms of $h_1,h_2,\cdots, h_{n-1}$. In fact it is a homogeneous polynomial of $h_j$'s of degree $2(n-2)$. For example in the cases of $n=3,4,5$
\begin{eqnarray*}
D_2&=&h_1^2-4h_0h_2\\
D_3&=&h_1^2h_2^2-4h_0h_2^3-4h_1^3h_3+18h_0h_1h_2h_3-27h_0^2h_3^2\\
D_4&=&h_1^2h_2^2h_3^2-4h_1^2h_2^3h_4-4h_1^3h_3^3+18h_1^3h_2h_3h_4\\
&&-27h_1^4h_4^2-4h_0h_2^3h_3^2+18h_0h_1h_2h_3^3
+16h_0h_2^4h_4\\
&&-80h_0h_1h_2^2h_3h_4+144h_0h_1^2h_2h_4^2-6h_0h_1^2h_3^2h_4\\
&&+144h_0^2h_2h_3^2h_4-128h_0^2h_2^2h_4^2-192h_0^2h_1h_3h_4^2\\
&&-27h_0^2h_3^4+256h_0^3h_4^3.
\end{eqnarray*}

On the other hand $h_j$'s are related with $x_j$'s according to
$$
x_k=h_k+Wh_{k-1},\qquad k=1,2,\cdots, n-1
$$
or, equivalently,
$$
h_k=x_k-x_{k-1}W+\cdots +(-1)^kW^k,\quad k=1,2,\cdots, n-1.
$$
Therefore the discriminant $D_{n-1}$ is a polynomial of $W$ of degree $(n-2)(n-1)$.

If we susbstitute $D_{n-1}(W)$ into $F$ of (\ref{t=int dW/F}), we find$$
t=\int^W{dW\over \sqrt{D_{n-1}(W)}}.
$$
The other variables $X_1, X_2,\cdots, X_{n-1}$ will be obtained similarly. Hence the orbits derived from the Nambu equations, whose Hamiltonians are elememtary symmetric polynomials, are given by hyper-elliptic functions. 

In the case of $n=3$ the motion of three variables $X_1, X_2, X_3$ are constrained on a circle fixed by the constants $x_1$ and $x_2$. We find
\begin{eqnarray}
X_1&=&{1\over 3}\left(x_1+2\sqrt{x_1^2-3x_2}\cos\left(\sqrt 3\ t\right)\right),\nonumber\\
X_2&=&{1\over 3}\left(x_1+2\sqrt{x_1^2-3x_2}\cos\left(\sqrt 3\ t-{2\pi\over 3}\right)\right),\nonumber\\
X_3&=&{1\over 3}\left(x_1+2\sqrt{x_1^2-3x_2}\cos\left(\sqrt 3\ t+{2\pi\over 3}\right)\right).\nonumber\\
\label{circle solution}
\end{eqnarray}

If all $x_j$'s are fixed besides $x_i$ the Nambu equation for $W=X_i$ becomes
$$
{dW\over dt}=(-W)^{n-i}\prod_{\stackrel{1\le k\le l\le n}{k,j\ne i}}(X_k-X_l)
$$
Note that the right hand side of this equation is just the square root of discriminant of the equation $P_{n-1,i}(X)=0$, where
$$
P_{n-1,i}(X)={\prod_{j=1}^n(X-X_j)\over X-X_i}
$$
The discriminant $D_{n-1,i}$ can be expressed in terms of $h'_1,h'_2,\cdots,h'_{n-1}$, where $h'_j$ is an elementary symmetric polynomial without $X_i$. As before,  $D_{n-1,i}$ is also a polynomial of $W$ of degree $(n-1)(n-2)$. We find
$$
t=(-1)^{n-i}\int^W{dW\over W^n\sqrt{D_{n-1,i}(W)}}.
$$

In the case of $n=3$ we could leave $x_2$ free, instead of $x_3$. Under the constraints
\begin{eqnarray*}
x_1&=&X_1+X_2+X_3\\
x_3&=&X_1X_2X_3
\end{eqnarray*}
being constant, we find an elliptic curve parameterized by
\begin{eqnarray*}
X_1&=&{\alpha\gamma\ {\rm sn}^2(u,k)\over \gamma-\alpha\ {\rm cn}^2(u,k)}\\
X_{2,3}&=&x_1-{\alpha\gamma\ {\rm sn}^2(u,k)\over \gamma-\alpha\ {\rm cn}^2(u,k)}\\
&&\pm{(\alpha-\gamma)^{3/2}\sqrt{\beta}\ {\rm cn}(u,k)\ {\rm dn}(u,k)\over 2(\gamma-\alpha\ {\rm cn}^2(u,k))\ {\rm sn}(u,k)}
\end{eqnarray*}
where 
$$
u={1\over 2}\sqrt{(\alpha-\gamma)\beta}\ t,\qquad k=\sqrt{{\alpha(\beta-\gamma)\over\beta(\alpha-\gamma)}}
$$
and $\alpha, \beta$ and $\gamma$ are the roots of 
$$
x^3-2x_1x^2+x_1^2x-4x_3=0.
$$

\subsection{$n$ dimensional generalization of Euler top and Nahm equation}

An $n$ dimensional box has $n(n-1)/2$ rectangles which are orthogonal with each other. Among them we choose $n$ independent rectangles. If we fix a relation between edge lengths of each of the $n$ rectangles, all $X_j$'s are determined, hence the box becomes rigid.

For example we can fix the diagonals of $n$ rectangles as follows:
\begin{equation}
x_j={1\over 2}\left(X_j^2+X_{j+1}^2\right),\qquad j=1,2,\cdots, n,
\label{diagonal}
\end{equation}
with $X_{n+1}=X_1$ to make the box rigid. The Jacobian of this map is, when $n$ is odd,
$$
\det J=(2X_1X_2\cdots X_n)^{-1}. 
$$

If $x_n=(X_n^2+X_1^2)/2$ is varied, all $X_j$'s will be changed simultaneously and $\mbox{\boldmath$X$}$ draws a curve in $\mbox{\boldmath$R$}^n$. The corresponding Nambu equations are
\begin{equation}
{dX_j\over dt}=(-1)^{n-j}{X_1X_2\cdots X_n\over X_j},\quad j=1,2,\cdots n.
\label{General Nahm}
\end{equation}
We can solve the constraints (\ref{diagonal}) for $X_j$'s 
$$
X_j^2=\alpha_j+(-1)^{n-j}W^2,\quad j=1,2,\cdots n-1
$$
where
$$
\alpha_j=2(x_j-x_{j+1}+\cdots -(-1)^{n-j}x_{n-1}).
$$
The right hand side of
$$
{dW\over dt}=X_1X_2\cdots X_{n-1}
$$
is given by a function of $W$. In fact we obtain 
$$
t=\int^W{dW\over\sqrt{\prod_{j=1}^{n-1}(\alpha_j+(-1)^{n-j}W^2)}}.
$$

Thus we conclude that the point $\mbox{\boldmath$X$}$ moves along a hyper-elliptic curve. When $n=3$, the solutions are given by the Jacobi elliptic functions as
\begin{eqnarray*}
X_1&=&\sqrt{2(x_1-x_2)}\ {\rm dn}(u,k),\\ 
X_2&=&\sqrt{2x_2}\ {\rm cn}(u,k),\\
X_3&=&\sqrt{2x_2}\ {\rm sn}(u,k),
\end{eqnarray*}
where
$$
u:=\sqrt{2(x_1-x_2)}\ t,\qquad k:=\sqrt{{x_2\over x_2-x_1}}.
$$

The above example can be readily generalized to the cases whose constraints can be reduced into the form
\begin{equation}
x_j={1\over 2}\sum_{k=1}^n\alpha_{jk}X_k^2,\quad j=1,2,\cdots, n-1.
\label{constraints 3}
\end{equation}
The Nambu equations are
$$
{dX_j\over dt}=(-1)^{n-j}\det A_j\ {X_1X_2\cdots X_n\over X_j},\quad j=1,2,\cdots,n
$$
where the matrix $A_j$ is given by eliminating the $j$th column from the $(n-1)\times n$ matrix $\{\alpha_{jk}\}$. By solving (\ref{constraints 3}) for $X_k$ as a function of $W=X_n$ and the constants and substituting them into
$$
{dW\over dt}=\det A_n\ X_1X_2\cdots X_{n-1}
$$
we again obtain a hyper-elliptic integral.

A simple case, {\it i.e.},
$$
A=\left(\matrix{1&-1&0\cr 0&1&-1\cr}\right),
$$
which is called the Nahm equation, was discussed in\cite{Takhtajan}. Another example is the famous Euler top corresponding to the matrix
$$
\left(\matrix{1&1&0\cr 0&1&1\cr}\right),
$$
which was discussed by Nambu\cite{Nambu}. We note that our generalization of this top to $n$ dimension is different from either one of \cite{Manakov} or \cite{Fairlie}.
\section{Completely integrable maps}

The hyper-elliptic functions have been known to solve soliton equations\cite{KM, DT}. They appear through the variation of sub-spectral parameters of Lax operators. We are going to show, in this section, that the hyper-elliptic solutions of soliton equations can be obtained equally by solving the Nambu equations, which we discussed in the previous section.

\subsection{Brief review of 3 point Toda lattice}

Before going into details of the discussion we will review briefly how the hyper-elliptic solutions are derived from soliton equations. In order to make clear the point of our arguments we consider a simple example, {\it i.e.}, 3-point Toda lattice. The time evolution of the 6-dynamical variables $(a_1,a_2,a_3,b_1,b_2,b_3)$ are determined by means of the Lax equation
\begin{equation}
{dL\over dt}=[B, L]
\label{Lax}
\end{equation}
where
$$
L=\left(\matrix{b_1&a_1&a_3\cr a_1&b_2&a_2\cr a_3&a_2&b_3}\right),\quad B=\left(\matrix{0&-a_1&a_3\cr a_1&0&-a_2\cr -a_3&a_2&0}\right).
$$
In addition to the periodicity condition, which imposes to $a_1, a_2,a_3$  a constraint 
\begin{equation}
a_1a_2a_3=1/8,
\label{aaa=1/8}
\end{equation}
the eigenvalues $\lambda_1,\lambda_2,\lambda_3$ of $L$ are constants of motion. Therefore only two variables out of six remain independent. It is conventional to choose two eigenvalues $\mu_1,\mu_2$ of the matrix $\left(\matrix{b_1&a_1\cr a_1&b_2\cr}\right)$ as new such variables. They are called subspectral parameters. The relations between the dynamical variables $(a_1,a_2,a_3,b_1,b_2,b_3)$ and $(\lambda_1,\lambda_2,\lambda_3,\mu_1,\mu_2)$ are algebraic. For instance we have
\begin{eqnarray*}
\mu_1+\mu_2&=&\lambda_1+\lambda_2+\lambda_3-b_3,\\
\mu_1\mu_2&=&b_1b_2-a_1^2
\end{eqnarray*}
Hence the time dependence of the dynamical variables can be found if we know how $\mu_1$ and $\mu_2$ vary in time.

Solving these algebraic relations for $(\lambda_1,\lambda_2,\lambda_3,\mu_1,\mu_2)$, the Lax equation (\ref{Lax}) can be converted into equations which determine the time evolution of $\mu_1,\mu_2$ as
$$
{d\mu_j\over dt}={1\over  4}{\sqrt{\Delta^2(\mu_j)-4}\over \mu_2-\mu_1},\quad j=1,2.
$$
Here $\Delta(\lambda)$ is a 3rd order polynomial of $\lambda$. Solutions to these equations are given in terms of elliptic functions.

We can summarize this result as follows. For the five dynamical variables, which are constrained by three constants of motion, we introduce two intermediate variables, which are also related with the dynamical variables algebraically. If we can find the dependence of the new variables on time, the time dependence of the five dynamical variables will be found by solving the five algebraic relations.

\subsection{Generalization to integrable maps}

Let us generalize this idea of solving 3-point Toda lattice to study larger class of integrable systems. For this purpose we consider a map $M(t)\rightarrow M(t+1)$ of an $m\times m$ matrix given by
\begin{equation}
M(t+1)=U^{-1}M(t)U.
\label{M=UMU}
\end{equation}

A large number of integrable maps have been known being represented in this form\cite{HT}. For an illustration we present here the discrete time $m$ point Toda lattice.
$$
M(t)=\qquad\qquad\qquad\qquad\qquad\qquad\qquad\qquad\qquad\qquad
$$
\begin{equation}
\left(\matrix{i_1+v_1&1&0&\cdots&0&i_1v_m\cr
                   i_2v_1&i_2+v_2&1&0&&0\cr
				   0&i_3v_2&i_3+v_3&1&&\cr
				   \vdots&&\cdots&&\vdots&\cr
				   0&0&\cdots&&&1\cr
				   1&0&\cdots&0&i_mv_{m-1}&i_m+v_{m}\cr}\right)
\label{M}
\end{equation}
$$
U(t)=
\left(\matrix{i_1&1&0&\cdots&0\cr 0&i_2&1&&0\cr
&&\vdots&&\cr 0&&&i_{m-1}&1\cr
1&0&\cdots&0&i_m\cr}\right)
$$
In the continuum limit of time the variables $(v_j, i_j)$ are related with $(a_j, b_j)$ of the Lax form by
$$
(v_j, i_j)=(2a_j,1-b_j)\qquad j=1,2,\cdots m.
$$

Suppose elements of the matrix $M$ in (\ref{M=UMU}) are determined in terms of $m+n-1$ dynamical variables with $n<m+1$. Since eigenvalues of the matrix, which we denote $\lambda_1,\cdots, \lambda_m$, are constant under the map, $m$ variables can be eliminated by solving algebraic relations between the elements of $M$ and the eigenvalues. The problem of solving the evolution equation (\ref{M=UMU}) is turned to finding proper intermediate $n$ variables. They must be responsible faithfully to a variation of the system under the constraints that the eigenvalues are conserved. We can use the Nambu equations to describe such a system.

In order to make concrete our argument we adopt the elementary symmetric polynomials as $m$ independent constants of the map:
\begin{eqnarray}
x_1&=&\lambda_1+\lambda_2+\cdots +\lambda_m\nonumber\\
x_2&=&\lambda_1\lambda_2+\lambda_1\lambda_3+\cdots +\lambda_{m-1}\lambda_m\nonumber\\
&\vdots&\nonumber\\
x_j&=&\sum_{1\le k_1<k_2<\cdots<k_j\le m}\lambda_{k_1}\lambda_{k_2}\cdots\lambda_{k_j}
\label{x=lambda}
\\
&\vdots&\nonumber\\
x_{m}&=&\lambda_1\lambda_2\cdots\lambda_m\nonumber
\end{eqnarray}
Writing them explicitly in terms of the elements $M_{jk}$ of the matrix $M$, we have
\begin{eqnarray}
x_1&=&M_{11}+M_{22}+\cdots +M_{mm}\nonumber\\
x_2&=& \sum_{1\le j<k\le m}(M_{jj}M_{kk}-M_{jk}M_{kj})
\label{x=aaa}
\\
&\vdots&\nonumber\\
x_m&=&\det M.\nonumber
\end{eqnarray}

Now we let $X_1,X_2,\cdots,X_n$ be the new $n$ intermediate variables and $x_1,x_2,\cdots, x_{n-1}$ be $n-1$ Hamiltonians of the system such that the intermediate variables are constrained by
\begin{eqnarray}
X_1+X_2+\cdots +X_n&=& x_1\nonumber\\
X_1X_2+X_1X_3+\cdots +X_{n-1}X_n&=&x_2\nonumber\\
&\vdots&
\label{XXX=x}\\
\sum_{1\le k_1<k_2<\cdots<k_{n-1}\le n}X_{k_1}X_{k_2}\cdots X_{k_{n-1}}&=&x_{n-1}\nonumber
\end{eqnarray}
The Nambu equations for the new variables are nothing but (\ref{dX_j/dt}) with $H_j=x_j,\ j=1,2,\cdots,n-1$ and solutions have been already discussed in IIIA. 
 
In order to find the behaviour of the matrix $M$ of (\ref{M=UMU}), we first identify (\ref{x=lambda}) and (\ref{x=aaa}) to express $m$ variables of the matrix in terms of the $m$ constants $\lambda_1,\cdots, \lambda_m$. The rest of the $n-1$ independent variables of the matrix $M$ will be determined by $X_1,X_2,\cdots,X_n$ if we identity $x_1,x_2, \cdots, x_{n-1}$ of (\ref{XXX=x}) with those of (\ref{x=aaa}). We would like to emphasize here that these steps will be done by purely algebraic procedures.

To be specific we consider the map (\ref{M=UMU}) with $M$ given by (\ref{M}). We further restrict to the case of $m=3$, {\it i.e,} the 3-point Toda lattice. Corresponding to the condition (\ref{aaa=1/8}) we may impose a constraint
\begin{equation}
v_1v_2v_3={\rm const.},
\label{vvv=const}
\end{equation}
so that the number of independent dynamical variables is five. Via explicit calculations we have
\begin{eqnarray}
x_1&=&i_1+i_2+i_3+v_1+v_2+v_3\nonumber\\
x_2&=&i_1 i_2+ i_1 i_3+i_2 i_3+i_1 v_2+ i_2 v_3+i_3 v_1\nonumber\\
&&\qquad\qquad+v_1 v_2+ v_1 v_3+v_2 v_3\nonumber\\
x_3&=&\left(1+i_1i_2i_3\right)\left(1+v_1v_2v_3\right)
\label{x=i,v}
\end{eqnarray}
in the place of (\ref{x=aaa}). The correspondence between (\ref{x=lambda}) and (\ref{x=aaa}) enables us to write three variables of $M$, say $i_1,i_2,i_3$, in terms of $v_1,v_2,v_3$ and the constants $\lambda_1,\lambda_2,\lambda_3$.

If we further identify $x_1,x_2$ in (\ref{x=i,v}) with those of (\ref{3 symmetric poly}), they, together with the condition (\ref{vvv=const}), determine $v_1,v_2,v_3$ as functions of $X_1,X_2,X_3$. Since $X_1,X_2,X_3$ have been known as given in (\ref{circle solution}), the behaviour of the matrix $M$ is determined.

\section{Remarks and discussions}

We have developed a method to derive Nambu equations from a given map (\ref{map}). There exist some ambiguities how to relate the time variable $t$ of the Nambu equations to the initial variables of the map. It could be one of the initial variables of the map as in the case of Proposition 1, or a function of it as it was the case of Proposition 2. They are not independent but are related with each other through a reparametrization of the variable $t$.

If the functions $f_1,f_2,\cdots,f_n$ of the map defined by (\ref{relations}) are purely algebraic, the map will determine an algebraic manifold. To study the nature of the manifold we change one of the initial variables of the map and see the response. Our propositions claim that the Nambu equations provide a systematic method to investigate such a response. By means of some examples we have shown that the manifolds described by certain types of map are characterized by hyper-elliptic curves.

When there are known some number of invariants under  time evolution, the Nambu equations determine the change of the dynamical variables. Since any function of the invariants is again an invariant there are many possible sets of Nambu equations. Suppose we can choose a proper set of invariants such that the functions are algebraic and the Nambu equations can be solved explicitly. Then the problem of solving the equations of motion are replaced to solve the algebraic relations among variables. We have demonstrated that the hyper-elliptic solutions of soliton equations can be derived in this way.


\end{document}